\documentclass[aps,prb,reprint,groupedaddress]{revtex4-1}

\usepackage{graphics}
\usepackage{epsfig}
\usepackage{amsmath}

\begin{document}

\preprint{}

\title{Measurement of the vortex mass in a superconducting film.}

\author{Daniel Golubchik}
\email[danielg@tx.technion.ac.il]{}

\author{Emil Polturak}

\author{Gad Koren}
\affiliation{Department of Physics, Technion - Israel Institute of Technology,
Haifa 32000, Israel}

\date{\today}

\begin{abstract}
We have combined high resolution magneto-optical imaging with an ultra-fast
heating/cooling technique to measure the movement of individual
vortices in a superconducting film. The motion took place while
the film was heated close to $T_c$, where pinning and viscous
forces are relatively small. Under these conditions, vortices move due to the magnetic repulsion between them. We found that a finite vortex mass has to be included in the analysis in order to account for the experimental results. The extent of the motion is consistent with a vortex mass being 3 orders of magnitude smaller than the mass of all the electrons in the core.
\end{abstract}

\keywords{Vortices, vortex mass, magneto-optical imaging}

\maketitle
Vortices in a superconductor are localized, topologically
protected excitations carrying a quantized amount of magnetic
flux. They are the "elementary particles" in several
models of statistical mechanics(Kosterlitz-Thouless, XY model). One
important parameter characterizing a particle is its mass. The
size of the mass decides whether vortices are quantum or classical objects\cite{Volovik:1997}. Quantum vortices in unconventional superconductors are predicted to support Majorana excitations\cite{Volovik:1999} and can be applied for quantum computing\cite{Kitaev:1999}. A quantized  vortex has no intrinsic mass. An effective mass however can be associated with its motion. Due to its importance, the concept of the vortex mass was discussed extensively over the years but remains a controversial issue. One point of view is that mass plays no role in the dynamics since an inertial term in the equation of motion of a vortex is always negligible next to the viscous drag force. Predictions for the value of the mass in different limits (dirty and superclean) span 5 orders of magnitude\cite{Suhl:1965,Chudnovsky:2003,Han:2005,Kopnin:1998}. Experimentally, the presence of an inertial term is difficult to detect since at low temperatures vortices in
superconductors are pinned, and if they move at all, their motion is dominated by viscosity. We are aware of only one such attempt, with inconclusive results\cite{Fil:2007}. It is quite clear that in order to check if an inertial term plays a significant role in vortex dynamics, one should realize conditions where the viscous drag force is not dominant.

Both pinning  \cite{Field:2002} and viscous drag forces
\cite{Larkin:1976} decrease strongly with temperature and vanish at the critical temperature $T_c$. In a thin superconducting film, the vortex-vortex
interaction ($\textbf{F}_{int}$)is mainly due to
the magnetic field of the vortices outside the
film\cite{Brandt:2009}. $\textbf{F}_{int}$ depends on temperature rather weakly.  Consequently, a temperature interval exists near $T_c$ in which the viscous and pinning forces can be smaller than the vortex-vortex
interaction, and do not dominate the dynamics. Our experiment was designed to measure vortex motion in this temperature interval. To obtain quantitative estimates, we use a classical equation of motion of a vortex \cite{Han:2005, Ao:1999}. This equation is derived by integrating over the microscopic degrees of freedom, leaving only macroscopic forces:

\begin{equation}
\mu_v d \frac{d\textbf{\emph{v}}}{dt}=\textbf{F}_{int}+\textbf{F}_L-\nabla{U_{p}}-\eta \textbf{\emph{v}}
\end{equation}

Here $\mu_v$ is the vortex mass per unit length, $d$ is the film thickness, $\textbf{F}_{int}$ is the vortex-vortex interaction,
$\textbf{F}_{L} \propto \textbf{J} \times \bf{\Phi}_{0}$ is the
Lorentz force due to interaction with currents \textbf{J} ($\Phi_0$ is the flux quantum), $U_{p}$
is the pinning potential, and $\eta\textbf{\emph{v}}$ is the viscous drag force. Measurements \cite{Beelen:1967} show that in Nb, the Magnus force \cite{Sonin:1997} is much smaller then the viscous force and can be neglected.
In the absence of external currents and fields, the Lorentz force
results from currents associated with vortices trapped
in the sample. Our observations were performed in a region near the center of the
sample where the currents vanish and this force is small\cite{Zeldov:1994}.

 The vortex-vortex interaction\cite{Brandt:2009} $\textbf{F}_{int}$ is repulsive
for vortices of the same polarity.  In the limit $r\gg\lambda$ ($\lambda$ is the penetration depth), $\textbf{F}_{int}$ depends on the intervortex spacing $r$ as $F_{int}=\Phi_0^2/\mu_0 \pi r^2$. This relation holds even at $0.97T_c$, the maximal temperature where we use it. At this temperature $\lambda\simeq0.3 \mu m$ while the minimal $r\sim 1 \mu$m.

\begin{figure}
\includegraphics[width=3.2in]{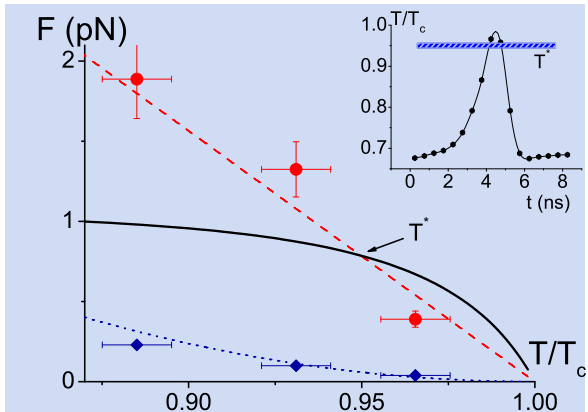}
\caption{Temperature dependence of the forces
acting on a vortex. The quantitative values are for a 200 nm Nb film. Solid black line is $\textbf{F}_{int}$
for two vortices separated by $1 \mu m$.
Circles (red) show the maximal drag force taken from our transport
measurements. Dashed line (red) is a linear fit. Blue diamonds and
dotted line represent the measured pinning force and the fit
to $F_p(T)=F_p(0)(1-T/T_c)^2$. $T^*$ is the crossover temperature
above which the pinning and the drag force are weaker than
$F_{int}$ and vortices are able to move. The inset shows the
time dependence of the temperature of the film during a heating pulse. Motion
of vortices takes place only during the short time interval where
the temperature exceeds $T^*$. \label{force}}
\end{figure}

We determined the pinning and the viscous drag forces for our films from transport measurements using the method described by Klein et al\cite{Klein:1985}. The temperature dependence of the various forces is shown in Fig.~\ref{force}. In the interval between $T^* \approx0.95T_c$ and $T_c$  $\textbf{F}_{int}$ is larger than the viscous drag force. Our experiment was performed in this interval.

The sample is a $200 nm$ thick Niobium film with $T_c$ of
$8.8K$ deposited on a sapphire substrate. The film is
patterned into squares of $400 \mu m \times 400 \mu m$. The experiment is shielded by $\mu$-metal from external magnetic fields. Small fields ($<$ mT) are applied using a solenoid
inside the shield. The vortices in the film are imaged using
high resolution magneto optics \cite{Golubchik:2009}. Using 10
s integration time, our system images large areas of $100 \times
100 \mu m^2$ with a $0.8\mu m$ resolution. Currently, our system
shows the highest resolution achieved with magneto optics.

Since the typical velocity of a vortex \cite{Bolz:2003} is few km/sec, it traverses the field of view in several nanoseconds. No technique can image this motion in real time.
Our approach is to take a pinned vortex array, release the pinning
for about 1 nsec, allowing the vortices to move under the various
forces, and then restore the pinning. Pinning can be turned off
and on by rapidly heating and cooling the film, as illustrated in the inset of Fig.~\ref{force}. During the short Infra-Red laser pulse, the temperature of the film is increased from the base temperature of 5.5K by an amount proportional to the intensity of the light. The $1 mm$
thick sapphire substrate is transparent at the wavelength of the
laser. Hence, by back side illumination, only the film heats up,
while the substrate remains near the base temperature. Heat
escapes from the film via ballistic phonons crossing into the cold
substrate. The substrate, which acts as a heat sink has a thermal
mass about $1000$ times larger than that of the film, and so its
temperature does not increases significantly during the pulse. The
thermalization time constant of the film is $\sim 6\cdot10^{-11}
sec$, so essentially the film remains hot as long as the laser
pulse is on, with its instantaneous temperature proportional to
the light intensity. We confirmed this scenario by comparing the
temperature profile measured directly using a $GeAu$ thin film
bolometer. The laser pulse intensity profile was measured simultaneously using a fast
photodiode. In our apparatus the length of the heating pulse is fixed, producing a temperature
profile shown in the inset of Fig.1. The static positions
of vortices are imaged before and after the motion
took place.

\begin{figure}
\includegraphics[width=3.2093in]{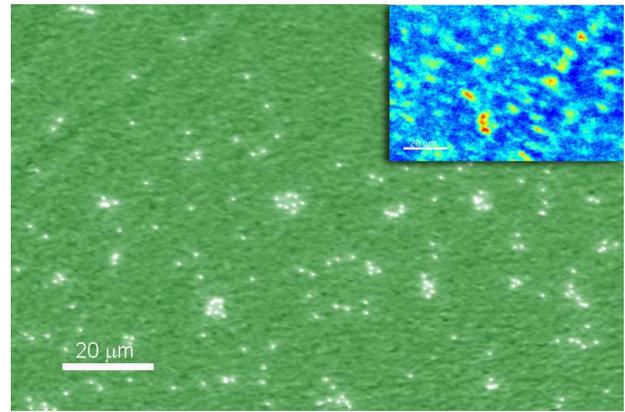}
\caption{Typical Magneto-Optical image of vortex aggregates in the
superconducting film. Each white spot is an individual vortex.
Inset: Light intensity distribution of a typical inhomogeneous
laser heating pulse. Red spots represent high intensity regions,
where the temperature during the pulse is above $T_c$.
\label{aggr}}
\end{figure}

Our initial vortex configuration contains small aggregates only few
vortices each, shown in Fig.~\ref{aggr}. To prepare such aggregates, we
first cool the film in a very small field, typically 30 $\mu$T.
The field is turned off at low temperature, leaving vortices trapped in the film. We then apply an inhomogeneous heating light pulse. The speckle pattern of the
light creates an inhomogeneous illumination, strong enough to
increase the temperature above $T_c$ in some small regions. Some of the
vortices escape into these normal regions and become trapped as
classical flux. Upon cooling, these regions become superconducting
again and the trapped flux disintegrates into small aggregates of
several vortices each. The typical distance between nearby
vortices in an aggregate is $\approx1 \mu m$. Typically, the aggregates are
separated by $\sim10 \mu m$, with the area in between largely free of
vortices.

\begin{figure}
\includegraphics[width=3.2093in]{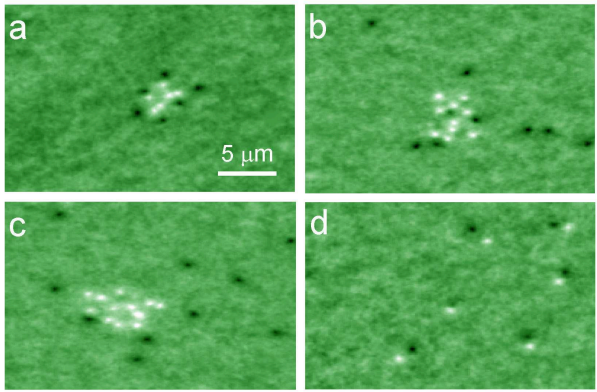}
\caption{Differential images of typical aggregates of
vortices before and after a homogeneous heating pulse. White and
dark spots mark the positions of the vortices before and after the
pulse. The scale bar for all the panels is the same. (a) an
aggregate heated to $T_{max}\simeq0.95T_C$. (b) an aggregate
heated to $T_{max}\simeq0.97T_C$. (c) an aggregate heated to
$T_{max}\simeq0.99T_C$ (d) isolated vortices heated to $T_{max}\simeq0.97T_C$.
\label{fig3}}
\end{figure}

We next apply a homogeneous heating pulse to the film containing the initial vortex array shown in Fig.~\ref{aggr}. To obtain uniform illumination (and heating), the light was passed
trough a diffuser. The random intensity variability across the
sample translated to temperature units was less then $0.01 T_c$. The displacement of vortices can be seen in the differential image shown in Fig.~\ref{fig3}, generated by subtracting the image before the pulse from the image after. If the maximal temperature during the pulse does not reach $\sim0.95T_c$, we see no motion within our resolution. Once the maximal temperature during the pulse exceeded $\sim0.95T_c$, the aggregates begin to disperse (shown in Fig.~\ref{fig3}a). This temperature is in good agreement with $T^*$ estimated using our transport measurements (see Fig.~\ref{force}). At $T=0.97T_c$, the typical displacement of the vortices is several $\mu m$  (see Fig.~\ref{fig3}b).  At $T=0.99T_c$, (Fig.~\ref{fig3}c) the displacement becomes comparable to the distance between aggregates. All the trapped vortices disappear at $T_c$.

An important check involves the displacement of isolated vortices. At $T_{max}=0.95T_c$ isolated vortices do not move. At $T_{max}=0.97T_c$ (Fig.~\ref{fig3}d) the displacement is significantly
smaller than that of vortices which are part of an aggregate
(Fig.~\ref{fig3}b). This confirms that within an aggregate the dominant force is
the vortex-vortex interaction ($\textbf{F}_{int}$), rather than
a force exerted by currents, which should affect isolated vortices in the same way as those in
aggregates. Along the same lines, once the vortices have dispersed, the intervortex repulsion $\textbf{F}_{int}$ decreases to the point where even if we heat the film again the vortices do not move.

When inspecting images taken at $T=0.97T_c$ (Fig.~\ref{fig3}b), we can
associate individual vortices at their final positions with a
particular initial aggregate, while at $T=0.99T_c$ this is not possible (Fig.~\ref{fig3}c). We therefore use only the data at $T=0.97T_c$ for quantitative analysis. Even within a specific aggregate, it is not straightforward to establish which of the vortices in the initial state is which in the final state. We therefore chose to evaluate the probability that a vortex is displaced by a distance $r$, assuming that the displacement is minimal. We start with a differential image of the whole field of view which typically includes $\sim$200 vortices. We then measure all the distances between each of the the vortices in the initial state and all those in a final state.
 From this data set, we take the shortest distance connecting two
vortices, one in the initial state and one in the final state.
These two vortices are then removed from the list and the process
is repeated. This step is done to avoid double counting. The results were averaged over 8 images with $\sim$1600 vortices in total. Dividing by the total number of vortices gives the probability density $\emph{P}_{exp}(r)$.  The algorithm works well as long as the the vortices do not
move further than a typical distance between the initial
aggregates which is true for data obtained  at
$T=0.97T_c$.

 To compare the experiment with theory, we generate images similar to Fig.~\ref{fig3} by numerical
integration of the equation of motion (Eq. 1). We start with initial configuration of typically 200 vortices taken from experiment (\emph{e.g.} fig.~\ref{aggr}). The integration was performed simultaneously on all the vortices. The time dependence of the temperature during the integration was taken from the experiment (shown in the insert of Fig.~\ref{force}).
At each step $F_{int}$ was calculated using the instantaneous positions of the vortices. Inside an aggregate, the initial value of $F_{int}$ is of the order of $10^{-12} $N. The pinning force at each temperature $F_{p}$ was taken from our transport measurements. $F_{p}$ shown in Fig.1, is consistent with previous measurements on Nb films\cite{Park:1992}. At $T\sim0.95T_c$, for our film
thickness $F_p \simeq6\cdot 10^{-14} N$, more than order of magnitude smaller than $F_{int}$.

The explicit form of the drag force $\eta \textbf{\emph{v}}$ is given by
\cite{Larkin:1976}
 \begin{equation}
\eta(v)\textbf{\emph{v}}=\eta(0)\frac{\textbf{\emph{v}}}{1+(v/v^*)^2}
\end{equation}
Here $\eta(0)$ is a sample
dependent parameter, and $v^*$ is a characteristic
velocity at which the drag force $\eta \textbf{\emph{v}}$ has a
maximum. The measured maximal drag force as a function of temperature is shown in Fig.~\ref{force}. Both $\eta(0)$ and $\emph{v}^*$ were determined for our films from our transport data. The values are consistent with prior measurements on Nb films \cite{Villard:2003}. Combining measured $\eta(0)$ and $\emph{v}^*$ with the instantaneous velocity $v$ allows us to determine the drag force at each integration step.

The sudden onset of motion at $T^*$ is consistent with the non linear form of the viscosity \cite{Larkin:1976}. Vortices are accelerated by $F_{int}$. As long as the velocity is below $v^*$ the drag force increases with $v$ and the motion is highly damped, limiting the distance traveled to less than 1 $\mu$m. At temperatures above $T^*$, $\eta \textbf{\emph{v}}^*$ becomes smaller than $F_{int}$. A vortex can be accelerated to $v>v^*$, experiences a reduced drag force, and travel distances significantly larger than 1 $\mu$m.

The accuracy of the simulations is limited by the experimental uncertainty in the initial positions of the vortices which set the exact value and direction of $F_{int}$. For that reason, the simulations can not reproduce exactly the experimental images.  We found however that the range of the motion of vortices which is important to define $\emph{P}^{sim}(r)$ is unaffected by this uncertainty.
To summarize, the pinning force turns our to be small, while the maximal viscous force $\eta \textbf{\emph{v}}^*$ and $F_{int}$ are comparable. We emphasize that with all the forces known, the vortex mass is the only free parameter in the calculation.

\begin{figure}
\includegraphics[width=3.2in]{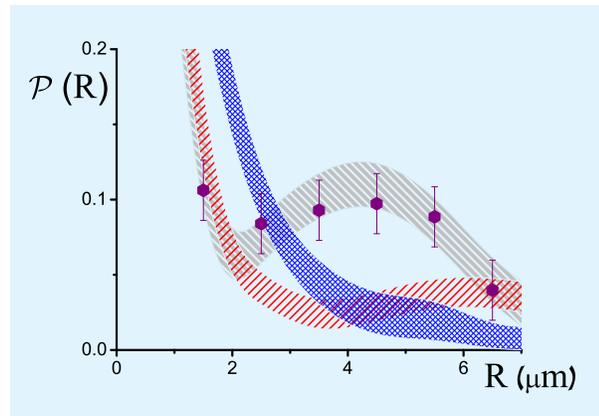}
\caption{Experimental (solid circles) and
calculated (bands) displacement probability distribution of
vortices at $T\simeq0.97T_c$.  Blue band corresponds to vortex masses of 20 $m_e/a$, gray band to 5 $m_e/a$, and red band  to 1 $m_e/a$. The high value of $\emph{P}(r)$ at small $r$ is largely due to isolated vortices which move less than 1 $\mu$m. The error bars and the width of the bands represent statistical uncertainty.} \label{disp_dist}
\end{figure}

Using the initial positions from the experiment and final positions from the simulations we calculated $\emph{P}_{sim}(r)$ by the same algorithm described above.
Similarly, $\emph{P}_{sim}(r)$ was averaged over 16
different images. The simulated and experimental $\emph{P}(r)$ are compared in Fig.~\ref{disp_dist}. The mass per unit length is given in units of electronic mass $m_e$ per lattice constant $a$. In simulations using large masses ($\mu_v>20 m_e/a$), the vortices move little ($r<2 \mu m$) and $\emph{P}_{sim}(r)$ is large only at small distances. For very small masses ($\mu_v<1 m_e/a$), the displacement can be large and $\emph{P}_{sim}(r)$ is significant at very large distances. Only for masses $5<\mu_v<10$, $\emph{P}_{sim}(r)$ is large for $2<r<5 \mu m$, which is in good agreement with experiment. This agreement suggests that in the temperature window close to $T_c$, Eq. 1 can successfully describe the dynamics of vortices. Furthermore, the results are sensitive to the presence of an inertial term. Taking into account all the experimental uncertainties, our final result is $1<\mu_v<20$.

Most of the literature quotes the vortex mass
at T=0. The first calculation of the vortex mass
by Suhl \cite{Suhl:1965} relied on the
velocity dependence of the free energy of the vortex core. This
gave a mass per unit length $\mu_v\sim1 m_e/a$. Additional contributions to the mass so
defined were discussed by several authors\cite{Chudnovsky:2003}.
Another approach is to define the mass through the proportionality
between the force acting on a vortex and its acceleration
\cite{Kopnin:1998,Volovik:1997,Han:2005}. The resulting mass is of
the order of the mass of all the electrons within the vortex core. To compare with our results, we need the value
close to $T_c$. Suhl used the Ginzburg-Landau
formalism\cite{Suhl:1965}, valid near $T_c$. For Nb  at
$T\simeq0.97T_c$ Suhl's mass is $\mu_v\sim10^{-2} m_e/a$. In the approach involving dynamic response
\cite{Kopnin:1998,Volovik:1997}, the temperature
dependence of the mass was calculated by
Han\cite{Han:2005}. At $T\simeq0.97T_c$ the mass is $\sim2\cdot
10^4 m_e/a$. Our result falls in between these values.
The only other reported measurement\cite{Fil:2007} comes
from the response of a vortex array to sound waves. In a large crystal, the vortices are tangled and the response of the array is collective. This may explain why the reported mass is 2 orders of
magnitude larger even than the largest theoretical prediction. In our experiment we use low
density vortices in a thin film. At low density, vortices are not entangled so the dynamic
response is that of a single vortex.

It was suggested that the vortex mass depends
on the "cleanliness" of the superconductor \cite{Sonin:1998}. Suhl's prediction corresponds to the dirty limit, while the dynamic response prediction to the superclean limit. Our sample is
intermediate between these limits (mean free path ($l\approx20$ nm) roughly equals the correlation length ($\xi_0\approx40$ nm)).
In conclusion, we have demonstrated that the inertial mass of a vortex is a meaningful concept which appreciably influences the dynamics close to $T_c$. "Quantum" vortices may indeed exist in the dirty limit of high $T_c$ superconductors.

We thank Ophir Auslaender, Eli Zeldov, Assa Auerbach and Edouard Sonin for
illuminating discussions, and Eyal Buks and Shmuel
Hoida for their contribution. This work was
supported by the Israel Science Foundation, and by
the Minerva and DIP projects.



\end{document}